\documentstyle[12pt,psfig,aaspp]{article}
\newcommand\degree{$^{\circ}$}

\newcommand\amin{\hbox{$^{\prime}$}}

\def\arcmin{\hbox{$^\prime$}}

\def\arcsec{\hbox{$^{\prime\prime}$}}

\def\utw{\smash{\rlap{\lower5pt\hbox{$\sim$}}}}

\def\udtw{\smash{\rlap{\lower6pt\hbox{$\approx$}}}}

\newcommand\hii{\mbox{H\hspace{0.2ex}{\sc ii}}}
\newcommand\hi{\mbox{H\hspace{0.2ex}{\sc i}}}

\newcommand\sii{\mbox{[{S\hspace{0.2ex}\sc ii}]}}

\newcommand\ha{\mbox{H$\alpha$}}

\newcommand\Sec{~sec$^{-1}$}

\newcommand{\EE}[2]{\mbox{{#1}~$\times$~10$^{#2}$}}
\newcommand{\TEN}[1]{\mbox{10$^{#1}$}}
\renewcommand\approx{\mbox{$\sim$}}

\newcommand\lt{$<$}
\newcommand\gt{$>$}

\newcommand\chisq{\mbox{$\chi^2$}}

\newcommand\ie{{i.e.}}
\newcommand\eg{{e.g.}}

\newcommand\etal{{et~al.}}

\newcommand{\Ref}{\par \hangindent=20pt \hangafter=1 \noindent}

\newcommand\acite[3]{{#1}~{#2}, {#3}}

\renewcommand\aj[2]{\acite{AJ}{#1}{#2}}
\newcommand\ana[2]{\acite{A\&A}{#1}{#2}}

\renewcommand\apj[2]{\acite{ApJ}{#1}{#2}}
\renewcommand\apjs[2]{\acite{ApJS}{#1}{#2}}
\renewcommand\apjl[2]{\acite{ApJ}{#1}{L#2}}

\renewcommand\mnras[2]{\acite{MNRAS}{#1}{#2}}

\renewcommand\pasj[2]{\acite{PASJ}{#1}{#2}}

\def \ampt      {\rlap{.}$'$}
\def \aspt      {\rlap{.}$''$}

\begin{document}

\title{X-rays from superbubbles in the Large Magellanic Cloud IV: \\ The
Blowout Structure of N44}

\author{Eugene~A.~Magnier,
You-Hua~Chu\footnote{
Visiting astronomer, Cerro Tololo Inter-American Observatory (CTIO),
National Optical Astronomy Observatories (NOAO), operated
by the Association of Universities for Research in Astronomy,
Inc. under cooperative agreement with the National Science
Foundation.},
Sean~D.~Points$^1$}
\affil{Astronomy Dept., University of Illinois, 1002 W. Green Street, Urbana,
IL 61801, USA; \\
gene@astro.uva.nl, chu@astro.uiuc.edu, points@astro.uiuc.edu}
\and
\author{Una~Hwang}
\affil{NASA/Goddard Space Flight Center, Code 666, Greenbelt, MD 20771, USA; \\
hwang@ferdi.gsfc.nasa.gov}
\and
\author{R.~Chris~Smith$^1$}
\affil{Dept. of Astronomy, University of Michigan, 834 Dennison, Ann Arbor, MI
48109, USA; \\
chris@astro.lsa.umich.edu}

\begin{abstract}
We have used optical echelle spectra along with ROSAT and ASCA X-ray
spectra to test the hypothesis that the southern portion of the N44
X-ray bright region is the result of a blowout structure.  Three
pieces of evidence now support this conclusion.  First, the
filamentary optical morphology corresponding with the location of the
X-ray bright South Bar suggests the blowout description (Chu \etal\
1993).  Second, optical echelle spectra show evidence of high velocity
(\approx 90~km\Sec) gas in the region of the blowout.  Third, X-ray
spectral fits show a lower temperature for the South Bar than the main
superbubble region of Shell~1.  Such a blowout can affect the
evolution of the superbubble and explain some of the discrepancy
discussed by Oey \& Massey (1995) between the observed shell diameter
and the diameter predicted on the basis of the stellar content and
Weaver et al.'s (1977) pressure-driven bubble model.
\end{abstract}

\clearpage

\section{Introduction}
Interactions between massive stars and the interstellar medium (ISM)
are among the dominant mechanisms involved in the chemical and
dynamical evolution of galaxies.  Massive stars deposit thermal and
kinetic energy into the ISM through their energetic winds and eventual
supernova explosions.  One incarnation of these interactions is the
``superbubble,'' a large (\approx 100-200 pc diameter), roughly
spherical interstellar shell around one or more OB associations.  A
superbubble shell consists of interstellar material swept up by the
stellar winds and supernova ejecta.  The shell interior contains hot
(10$^6$ K), shocked wind and ejecta with density of $\sim0.01$
cm$^{-3}$ surrounded by the cooler, denser gas of the shell (Castor,
McCray, \& Weaver 1975; Weaver et al. 1977).

A large number of superbubbles have been identified in Local Group
galaxies.  The Large Magellanic Cloud (LMC) is a nearly ideal
laboratory in which to study the detailed physics of these
interstellar structures, as the LMC has a low foreground interstellar
extinction and a small inclination angle of \approx 40\degree\ (Feast
1991), minimizing confusion along the line-of-sight.  At the LMC
distance of 50 kpc, 1\arcmin\ corresponds to 15 pc, allowing us to
study the gas portions of these structures in great detail with a
variety of instruments, as well as to resolve the bulk of the
individual massive stars with ground-based telescopes.  Dozens of
superbubbles are known in the LMC, giving many examples for study.

N44 is among the most luminous \hii\ complexes in the LMC, and
consists of three OB associations and several regions of optical
nebulosity (Henize 1956; Lucke \& Hodge 1970).  A variety of
\ha-bright structures are seen in N44, and identifications of the
components, both bright knots and large filamentary arcs, have been
made by Henize (1956) with labels N44 A -- N and Davies, Elliott, \&
Meaburn (1976) with labels DEM L140 -- 170.  Figure \ref{Halpha}a
shows an \ha\ CCD image of the region with the major emission regions
marked.  Meaburn \& Laspias (1991) detected the expansion of the two
of the large shells, DEM L140 and DEM L152, confirming their
description as superbubbles.

Diffuse X-ray emission from various portions of the N44 region was
detected by the {\em Einstein Observatory} (Chu \& Mac Low 1990; Wang
\& Helfand 1991).  Observations of the region with ROSAT (Chu et al.
1993, hereafter C93) allowed for a more detailed comparison between
the X-ray and H$\alpha$ morphologies, and showed several distinct
features.  C93 used the X-ray morphology to define five regions of
particular interest: Shells 1 -- 3, the South Bar, and the North Diffuse
region.  Among the three brightest regions, Shell 1 was identified as
an X-ray-bright superbubble, the South Bar was suggested to be a
``blowout structure'' from the superbubble, and Shell 3 was identified
as a previously unknown supernova remnant (SNR).  These three bright
X-ray features are marked in Figure 2a, a ROSAT Position Sensitive
Proportional Counter (PSPC) image overlaid with H$\alpha$ contours.
The North Diffuse region is outside of the bounds of this picture and
Shell 2 is not shown and will not be discussed further in this paper.
Shell 1 corresponds primarily with the interior of the optical shell
DEM L152.

The stellar content of DEM L152 has recently been discussed by Oey \&
Massey (1995 -- hereafter OM95), who make model comparisons between
the observed expansion velocities and the expected energy input from
the stellar winds.  Two important results are drawn from their study.
First, the stellar content interior to the shell is generally older
than the population immediately exterior to the shell ($\sim10$ Myr
vs.\ $<$5 Myr), suggestive of sequential star formation.  Second,
their superbubble evolution models, adopting stellar wind and
supernova energy inputs implied by realistic stellar contents,
over-predict the shell diameter.  Several possible reasons for this
size discrepancy are given by OM95; however, we believe that part of
the discrepancy may be caused by significant energy/pressure lost via
the blowout shown in C93.

In this paper, we have combined newly-obtained \ha\ CCD images,
high-resolution echelle spectroscopy and ASCA X-ray observations,
along with the existing ROSAT observations to test the hypothesis that
the South Bar is the result of a blowout.

\section{Optical Echelle Spectra}

We have used optical spectroscopy to explore the velocity field in the
vicinity of the proposed blowout region of the X-ray South Bar.  We
obtained high-dispersion spectra at the Cerro Tololo Inter-American
Observatory (CTIO) using the echelle spectrograph with the long-focus
red camera on the 4m telescope.  We used the instrument as a
single-order, long-slit spectrograph by inserting a post-slit
interference filter and replacing the cross-dispersing grating with a
flat mirror so that a single echelle order around H$\alpha$ and
[N\,II] $\lambda \lambda$6548~\AA, 6583~\AA\ could be imaged over the entire
slit.  The 79 l/mm grating was used for these observations, and the
data were recorded with a Tek 2048$\times$2048 CCD using Arcon 3.6
read-out.  With a pixel size of 24$\mu$m pixel$^{-1}$, this CCD
provided a sampling of \approx 0.082~\AA\ (3.75 km~s$^{-1}$) along the
dispersion and 0\arcsec.267 on the sky.  Spectral coverage was
effectively limited by the bandwidth of the interference filter to
125~\AA.  Coverage along the slit was 220\arcsec, including an
unvignetted field of \approx 2\arcmin.5  The instrumental profile, as
measured from Th-Ar calibration lamp lines, was 16.1 $\pm$ 0.8
km~s$^{-1}$ FWHM.  The ranges include variations in the mean focus and
variations in the focus over a single spectrum.  Spatial resolution
was determined by the seeing, which was \approx 1\arcsec.

The data were bias subtracted at the telescope.  Flat-fielding,
wavelength calibration, distortion correction, and cosmic ray removal
were performed using standard IRAF\footnote{IRAF is distributed by the
National Optical Astronomy Observatories (NOAO).} routines.  Wavelength
calibration and curvature correction were performed using Th-Ar lamp
exposures taken in the beginning of each night.  To eliminate velocity
errors due to flexure in the spectrograph, the absolute wavelength
scale in each spectrum was referenced to the telluric H$\alpha$
airglow feature.

Figure \ref{Halpha} shows the echelle spectrum along with our recent
\ha\ CCD image.  The CCD image was taken at the Curtis Schmidt
telescope at CTIO.  This image, along with several other filter
images, was taken on February 18, 1994, using the Thompson
1024$\times$1024 CCD.  With this setup, the CCD has 1\aspt{835}
pixels, giving a field of view of 31\ampt{3}.  Further details of the
Curtis Schmidt observations can be found in Smith \etal\ (1994).  The
location of the echelle slit used for the spectrum is indicated on the
CCD image.  The star labeled ``Star 1'' shows up clearly at the center
of the spectrum.  Several cuts in the velocity direction are plotted
below the spectrum, and the heliocentric velocity scale is given to
the left.  The most important feature in this spectrum is the
variation in the range of velocities seen.  The regions to the east of
Star 1 have only a small range of velocities, roughly $-30$ to $+40$
km\Sec, while to the west, there is clear evidence of significantly
blue-shifted material, with velocity offsets up to $\sim -90$ km
s$^{-1}$.  The region with this high-velocity feature coincides with
the brightest part of the X-ray South Bar.  Also in this area, optical
filaments can be seen which extend away from the main optical shell of
DEM L152.  The high-velocity feature is not detected elsewhere in N44
(Meaburn \& Laspias 1991; Hunter 1994), except at another position
within the X-ray South Bar (Figure 15 of Hunter 1994).  Therefore, the
high-velocity gas lends further support to the blowout nature of this
region.

\section{X-ray Observations}

If the South Bar is indeed a blowout structure, with the gas escaping
into a region of lower pressure beyond the superbubble shell, it is
expected that the temperature of the hot escaping gas should drop as
it undergoes adiabatic expansion.  To test this hypothesis, we have
used X-ray observations from both the ROSAT and ASCA X-ray
observatories to compare the temperatures in Shell 1 and the South
Bar.

The ROSAT PSPC observations have been discussed in some detail in C93.
We present here a brief summary: The observations were performed March
7 -- 9, 1992, and are archived under observation number RP500093.  For
these observations, N44 was centered in the PSPC, where the angular
resolution was \approx 30\arcsec.  The PSPC is sensitive in the energy
range of 0.1 -- 2.4 keV and has an energy resolution of \approx 43\% at
1 keV.  The total effective exposure time was 8871 seconds.

Our ASCA observations were performed March 1, 1995.  The ASCA
observatory has been discussed by Tanaka et al. (1994).  Briefly, the
satellite carries four imaging thin-foil grazing incidence X-ray
telescopes.  Two of the telescopes are focused on solid-state CCD
imaging spectrometers, called SIS\,0 and SIS\,1.  The other two
telescopes use gas imaging spectrometers (GIS) as the detectors.  In
this paper, we have only used the SIS data because of the low
sensitivity of the GIS for energies below \approx 1 keV, where our
objects are the brightest, and the substantially lower energy
resolution compared with the SIS.  The SIS detectors have an energy
resolution of \approx 2\% at 6 keV and cover the energy range 0.4 --
10 keV, with reduced throughput near the ends of the band.  The
spatial resolution is poor compared to the ROSAT PSPC, with a narrow
core of \approx 1\amin\ diameter and a half-power radius of 3\amin.
Both of the SIS detectors are composed of 4 separate CCDs.  Not all
CCDs must be run for a given observation, and the presence of ``hot''
and ``flickering'' pixels, which fill the telemetry with false
signals, have made it necessary to use only a subset of each detector
for many observations.

  Because of the telemetry limitations, our observations were
performed in two-CCD mode, with the South Bar placed at the standard
point source location, near the center of the focal plane on SIS0,
chip 1 (SIS1 chip 3; see Figure \ref{Xray}b).  The observations were
performed in faint mode when allowed by the telemetry, and converted
to bright mode on the ground for analysis.  Using the standard
processing software ({\em ascascreen} \& {\em ftools}), we removed bad
time periods using the following selection criteria: aspect deviation
\lt 0.01\degree, angle to bright earth \gt 20\degree, elevation \gt
10\degree, minimum cutoff rigidity of 6, PIXL rejection of 75.  We
also removed hot and flickering pixels.  After screening, a total of
32,964 sec of usable data remained.  We included high, medium, and low
bit rate data.  Although low bit rate data have been seen to have
difficulties in the past, we did not reject them as they constituted \lt
1\% of our total integration time.

We extracted spectra from the ROSAT and ASCA data using regions which
were designed to be as similar as possible to those used in C93.  The
background was chosen in a different location because the region used
by C93 lies outside of the ASCA SIS field of view.  The background
regions were chosen to exclude point sources and obvious diffuse
emission which would contaminate the spectra.  The background was taken
from a source-free region on chip 0 of SIS0 (chip 2 of SIS1) because
there was no region on chip 1 of SIS0 (chip 3 of SIS1) large enough to
make a useful measurement.  A comparison between a small source free
region on chip 1 of SIS0 (chip 3 of SIS1) showed no significant
difference with the background we have used.  It was necessary to use
a local background region in the field because much of the background
is contributed by diffuse emission from the LMC itself, in addition to
the usual sources of background, such as charged particles, scattered
solar light, and cosmic X-ray background.  A comparison with the
archive background fields shows significant excess emission in the
local background field: 0.029 cts\Sec\ for the local background vs
0.018 for the same region of the chip in the archive background field.

For the ASCA data, spectra were extracted from both SIS0 and SIS1.
The datasets from both SIS telescopes and the PSPC cannot be simply
combined directly (\ie, added together) as the three telescopes have
distinct response matrices.  Instead, the three spectra were fit
jointly to the same models using XSPEC.  The normalizations were
allowed to fit separately to allow for differences in spatial and
spectral resolution.  Spectral bins for all three instruments were add
together in order to give sufficient counts in each spectral bin that
the \chisq\ statistics would be valid.  Optically thin thermal spectra
from Raymond \& Smith (1977, hereafter RS) and Mewe-Kaastra (Mewe
\etal\ 1985; Kaastra 1992, hereafter MEKA) were used.  The abundances
of all elements except Fe were set to 30\%, consistent with the LMC
abundances.  Because of some discrepancies in certain Fe line ratios,
we allowed the Fe abundance to float to the best fit value, and fixed
it at that value to determine the temperature confidence contours.  In
all fits, the Fe abundances were in the range 1.5\% to 7.9\% solar,
improving the fit substantially over models with the Fe abundance set
to 30\%.  We also experimented with letting other elemental abundances
float, but no single element improved the fits substantially.  We
found that both RS and MEKA models with low Fe abundances were in
reasonable agreement with the observed spectra.  Derived parameters
from the two models were in good agreement with each other.

Figure~\ref{S1SB} presents the results of the spectral fits for Shell
1 and the South Bar using the MEKA model.  The confidence contours and
best-fit spectra are shown.  Figure~\ref{Shell3} shows the confidence
contours and best-fit spectra for Shell 3.  The contours suggest that
the expected temperature difference between Shell 1 and the South Bar
is present: The blowout region (South Bar) has a 99\% confidence range
for the temperature of 0.29 -- 0.49 keV, while the main superbubble
(Shell 1) has a 99\% confidence range of 0.49 -- 0.57 keV, with no
overlap of the 99\% confidence contours between the two fits.  The
supernova remnant proposed by C93, Shell 3, does not show a
significantly higher temperature as previously claimed -- the 99\%
confidence interval is 0.24 -- 0.46 keV.  However, this does not
refute the identification of this structure with an SNR, and is
perhaps more consistent with the observed size of the structure (30
pc) and correspondingly large estimated age (\approx \EE{1.8}{4} yr)
reported by C93.  The South Bar and Shell 1 have a low absorption,
\approx \TEN{21} cm$^{-2}$, consistent with the optical and radio
measurements of the hydrogen column density (see discussion in C93).
The best fit absorption for Shell 3 is \EE{3.8}{21} cm$^{-2}$.
Although the error contour for Shell 3 does not constrain the
extinction strongly (\EE{1.2 -- 6.5}{21} cm$^{-2}$), it is worthwhile
to note that the somewhat higher extinction of Shell 3, compared with
Shell 1 and the South Bar is not unreasonable given the presence of
faint dust lanes visible across Shell 3 in the \ha\ images.

While these measurements appear to show quite a strong difference in
the temperatures of the South Bar and Shell 1, we caution the reader
that the confidence contours do not completely describe the true
errors involved.  In particular the fact that, without allowing the Fe
abundance to float, the model spectra deviate systematically from the
observed spectra is a point for concern.  The problem of
poorly-fitting Fe lines has been noted already and blamed on errors in
the atomic physics (see \eg, Fabian et al. 1994; Liedahl et
al. 1995). Several groups are currently working on updated model
calculations incorporating newer atomic data in an effort to improve
the situation.  Nonetheless, even without such improved models, the
relative temperatures of the South Bar and Shell 1 determined with the
MEKA and RS models are quite believable.  One particular concern of
the fits presented here is that there can be a small correlation
between the Fe abundance for a fit and the temperature. To show the
effect of this correlation, which also plot confidence contours for
the fit to the South Bar, keeping the Fe abundance fixed to the best
fit value for Shell 1 (gray contours).  The significance of the
resulting temperature difference is reduced, but still present -- now
there is overlap between the 99\% contours, but the 90\% confidence
contours do not touch.

\section{Discussion}

We have used optical echelle spectra along with ROSAT and ASCA X-ray
spectra to test the hypothesis that the southern portion of the N44
X-ray bright region is the result of a blowout structure.  Three
pieces of evidence now support this conclusion.  First, the
filamentary optical morphology corresponding with the location of the
X-ray bright South Bar suggests the blowout description (C93).
Second, optical echelle spectra show evidence of high velocity
(\approx 90 km\Sec) clouds in the region of the blowout.  Third, X-ray
spectral fits show a lower temperature for the South Bar than the main
superbubble region of Shell 1.  If the South Bar were the result of
escaping gas expanding adiabatically into the surrounding region of
lower pressure gas, the temperature would be expected to drop, as is
observed.  The combination of the optical and X-ray morphology, the
high velocity gas seen in the echelle spectra, and these tentative
X-ray temperature determinations give strong support to the
interpretation of this structure of the South Bar as a blowout from
the main shell of the superbubble as suggested by C93.

Such a blowout may explain the discrepancies discussed by OM95 between
the observed shell diameter and the diameter predicted on the basis of
the stellar content and Weaver et al.'s (1977) pressure-driven bubble
model.  Below we will discuss N44's superbubble energetics and the
effects of the blowout.

First we use OM95's stellar content to estimate the total energy input
to Shell 1 of N44.  OM95 have derived an age of 10 Myr for the stars
within the superbubble and 5 Myr for the stars on the exterior.  These
ages will be used as references.  We have integrated the stellar wind
power implied by the stellar content (Figure 11 of OM95), and obtained
a total stellar wind energy input of 3.2$\times10^{51}$ erg over the
first 5 Myr, and 3.5$\times10^{51}$ erg over the first 10 Myr.
Apparently, the stellar wind input during the second 5 Myr is only
about 10\% that of the first 5 Myr.  The initial mass function of the
stars in N44 implies that 1 -- 4 supernovae have occurred in the past 10
Myr.  It is likely that supernovae dominate the energy input in the
second 5 Myr as each explosion deposits \approx \TEN{51} erg.  The
total energy input from the stellar winds and supernovae is probably
in the range of $(4-7)\times10^{51}$ erg.

We may use the model fits derived from the ASCA SIS and ROSAT PSPC
data to determine more accurately the thermal energy in the
superbubble interior.  Using a log N$_{\rm H}$ of 21.0 and a kT of
0.55 keV for the superbubble interior, the observed ROSAT flux gives a
normalization factor log (N$_e^2$V/4$\pi$D$^2$) = 11.05 in cgs units,
where N$_e$ is the electron density, V is the X-ray emitting volume,
and D is the distance to the LMC (50 kpc).  We use the ROSAT flux
since the better angular resolution of the PSPC implies that fewer
photons are lost due to high-angle scattering.  If we assume that the
X-ray emitting volume is only 1/2 of the superbubble volume
(\EE{1.2}{5} pc$^3$), we derive an N$_e$ of 0.14 cm$^{-3}$, a mass of
194 M$_\odot$, and a thermal energy of 2$\times$10$^{50}$ erg.  If the
X-ray emitting volume is only 1/100 of the superbubble volume, because
of a small filling fraction, then N$_e$ is increased to 1.0 cm$^{-3}$,
the mass is reduced to 28 M$_\odot$, and the thermal energy is reduced
to 3$\times$10$^{49}$ erg.

In Weaver et al.'s (1977) pressure-driven wind-blown bubble model, the
thermal energy in the hot interior should be about 5/11 of the total
stellar wind energy input.  We see clearly that the thermal energy
derived from our X-ray observations is at least an order of magnitude
lower than expected.  Assuming the input energy from stellar winds and
supernovae is correct, this energy discrepancy can be caused by two
effects -- energy leakage through the blowout and energy loss via
radiation.  To test the effect of energy loss via cooling, we note that
the cooling timescale is roughly (1/N$_e$) Myr.  For the possible
range of N$_e$ values, the cooling timescale would be a few Myr;
therefore, we may expect a reduction of the thermal energy by a factor
of 2, or a few at most.  Another mechanism must account for the rest
of the lost energy -- one possibility is the blowout.

The amount of energy lost in the blowout can be roughly estimated from
the X-ray data.  Using the ASCA SIS model fit of the blowout region,
log N$_{\rm H}$ = 21.0 and kT = 0.41 keV, we obtain a normalization
factor log (N$_e^2$ V / 4$\pi$D$^2$) = 10.85 in cgs units.  If we
assume that the depth of the X-ray emission is the same as the width,
the emitting volume is 4$\times$10$^4$ pc$^3$, N$_e$ is 0.13
cm$^{-3}$, the thermal energy is 1.0$\times10^{50}$ erg, and the mass
is 133~M$_\odot$.  Note that the X-ray emitting volume could be
overestimated by a factor of 2 or more, then the N$_e$ would be a
factor of $\sqrt{2}$ higher, and the thermal energy and the mass would
be a factor of $\sqrt{2}$ lower.  This amount of thermal energy, lost
from the superbubble, is similar in magnitude to the thermal energy
remaining in the superbubble.  The largest uncertainty in this
comparison is in the determinations of the emitting volume in the
superbubble interior and the blowout.  Future High Resolution Imager
images may help reduce the uncertainties.

We now examine the thermal energy conversion problem with the blowout
and radiative cooling taken into consideration.  If we use the low
estimates for the electron density, we obtain high estimates for the
thermal energies in the superbubble interior and the blowout,
2$\times$10$^{50}$ erg and 1$\times10^{50}$ erg, but also high
estimates for the radiative cooling timescale.  Applying the Weaver et
al. (1977) model, a total energy input of $(4-7)\times10^{51}$ erg
should have $(2-3)\times10^{51}$ erg converted to thermal energy in
hot gas.  The observed thermal energy in the superbubble interior and
the blowout together is only 3$\times$10$^{50}$ erg.  Even if we
assume that 1/2 of the thermal energy has been radiated away (thereby
ignoring the high estimate for the cooling timescale), there is still
a discrepancy by a factor of 3-5.  If we use the high estimates for
the density, we would have much lower estimates for the thermal
energies, and the discrepancy between observed and expected thermal
energies would be even larger.  Therefore, we conclude that either the
Weaver et al. (1977) model is drastically wrong or the stellar wind
energy input must have been over-estimated -- we consider the latter
to be more likely.  A similar conclusion has been reached by Garc\'\i
a-Segura \& Mac Low (1995) in their modeling of the single-star
wind-blown bubble NGC 6888; they suggest that stellar wind strengths
could have been over-estimated by a factor of $\ge3$ because stellar
winds are clumpy (Moffatt \& Robert 1994).

Besides the interior thermal energy, another important reservoir of
energy in a superbubble is the expansion of the cool \hii\ shell.  The
radius of N44 superbubble is 30 pc, and the expansion velocity is
$\sim$40 km s$^{-1}$ (Meaburn \& Laspias 1991).  The density in the
shell is unfortunately in the low-density regime for the \sii\
doublet, so it is necessary to follow the procedure outlined by Chu et
al. (1995) to derive a rms density from the peak emission measure
along the shell rim.  The shell thickness is calculated assuming an
isothermal shock into a photo-ionized interstellar medium.  For a peak
emission measure of 1500 -- 4000 cm$^{-6}$ pc along the rim of the shell
(C93), we derive a shell thickness of 0.5 pc, a rms electron density
of 12 -- 20 cm$^{-3}$ in the \hii\ shell, an ambient density of 0.6 -- 1.0
cm$^{-3}$, a shell mass of $1,600-2,800$ M$_\odot$, and a shell
kinetic energy of $(2.6-4.4)\times10^{49}$ erg.  This kinetic energy
is at least an order of magnitude lower than expected from Weaver et
al.'s model assuming a total energy input of $(4-7)\times10^{51}$ erg.
This discrepancy may also be caused by the energy leakage and radiation
losses, or by an over-estimate of the stellar wind strengths as
discussed above.

The presence of a significant blowout in N44, and the resulting effect
on the evolution of the superbubble, has important implications for
the hot phase of the ISM and model calculations of superbubbles in
galaxies.  It is interesting to note that the blowout in N44 is
probably within the gaseous disk, instead of into a halo.  This
assumption is based on the small size of N44 compared with the the
scale height of the \hi\ layer, as well as the apparent interaction
with the \hii\ regions to the south.  If such blowouts are common in
superbubbles in general, theoretical estimates of superbubble sizes
will be systematically too large.  Such an error would affect model
predictions of the ISM distribution in galaxies.  It is reasonable to
conclude that blowouts of this nature may be common: the blowout
discussed in this paper is not trivial to detect, and depended upon
the X-ray morphology information.  Since N44 is the brightest
superbubble in X-rays in the LMC, the blowout in this case is
relatively easy to detect.  For fainter superbubbles, such blowouts
could be common and may simply have gone unnoticed so far.  Future
work to search for evidence of other, similar blowout structures
should be performed.

\acknowledgements
We acknowledge the support of NASA grants NAG 5-1900, NAG 5-2679, NAG
5-2973, and NAGW-4519.  SDP acknowledges support by a National Science
Foundation graduate research fellowship.  Thanks to Joel Parker for
helping to start our collaboration.

\newpage
\section*{Figure Captions}

\begin{figure*}
\psfig{file=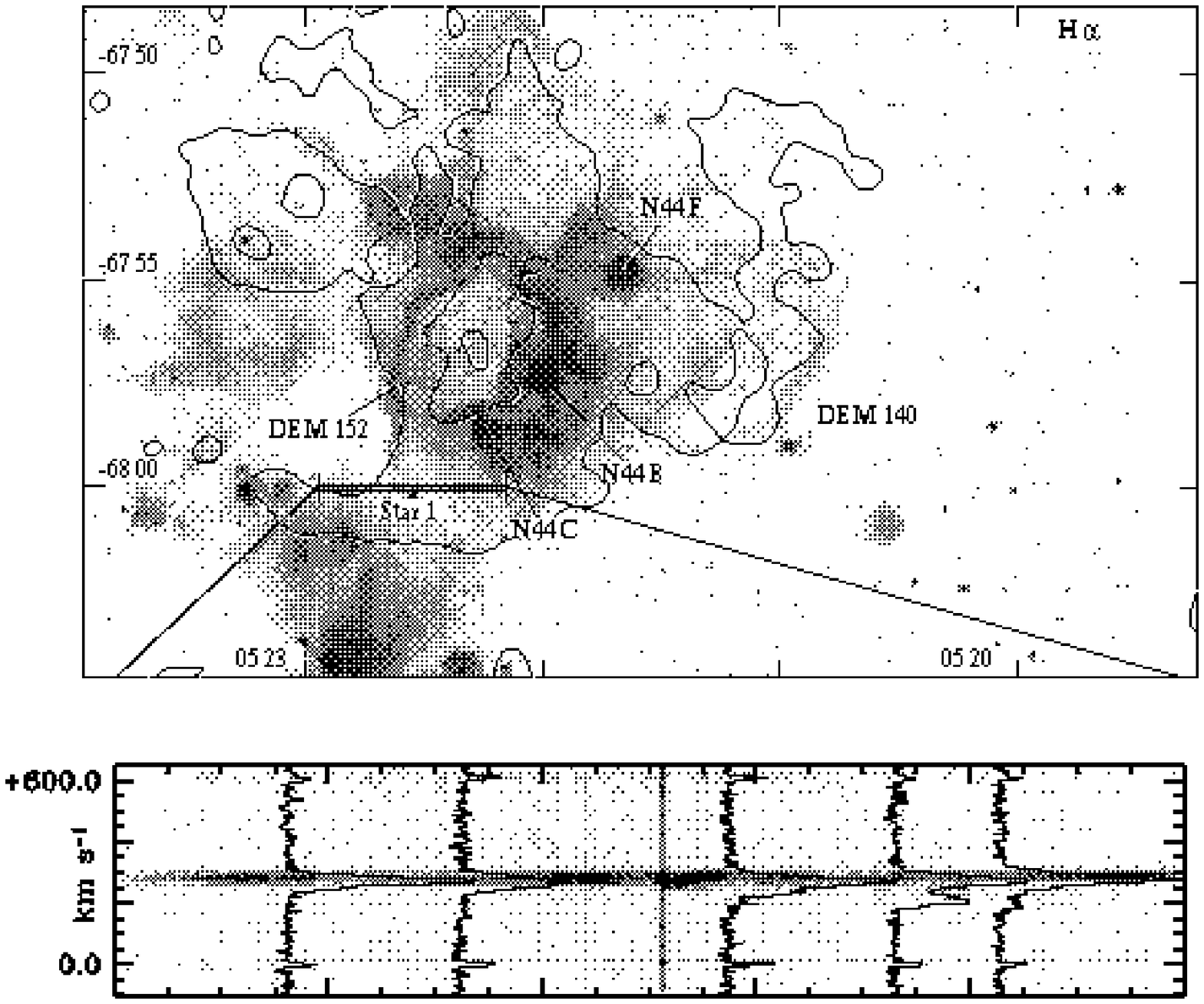,width=16.5cm,silent=}
\caption{\label{Halpha}  }
\end{figure*}

\noindent
Figure~\ref{Halpha}: top) \ha\ image of the N44 complex with ROSAT
X-ray contours overlaid.  The major regions of nebulosity are labeled
(Henize 1956; DEM).  The location of the echelle slit is noted, along
with the position of Star 1.  bottom) The echelle spectrum, including
five representative cuts in the spectral direction.  The velocity axis
is at the left, showing the heliocentric velocity.  In this figure,
the contours are draw in pairs, with a white line at a slightly lower
value than the black line. This allows the reader to distinguish
positive and negative contours.

\begin{figure*}
\psfig{file=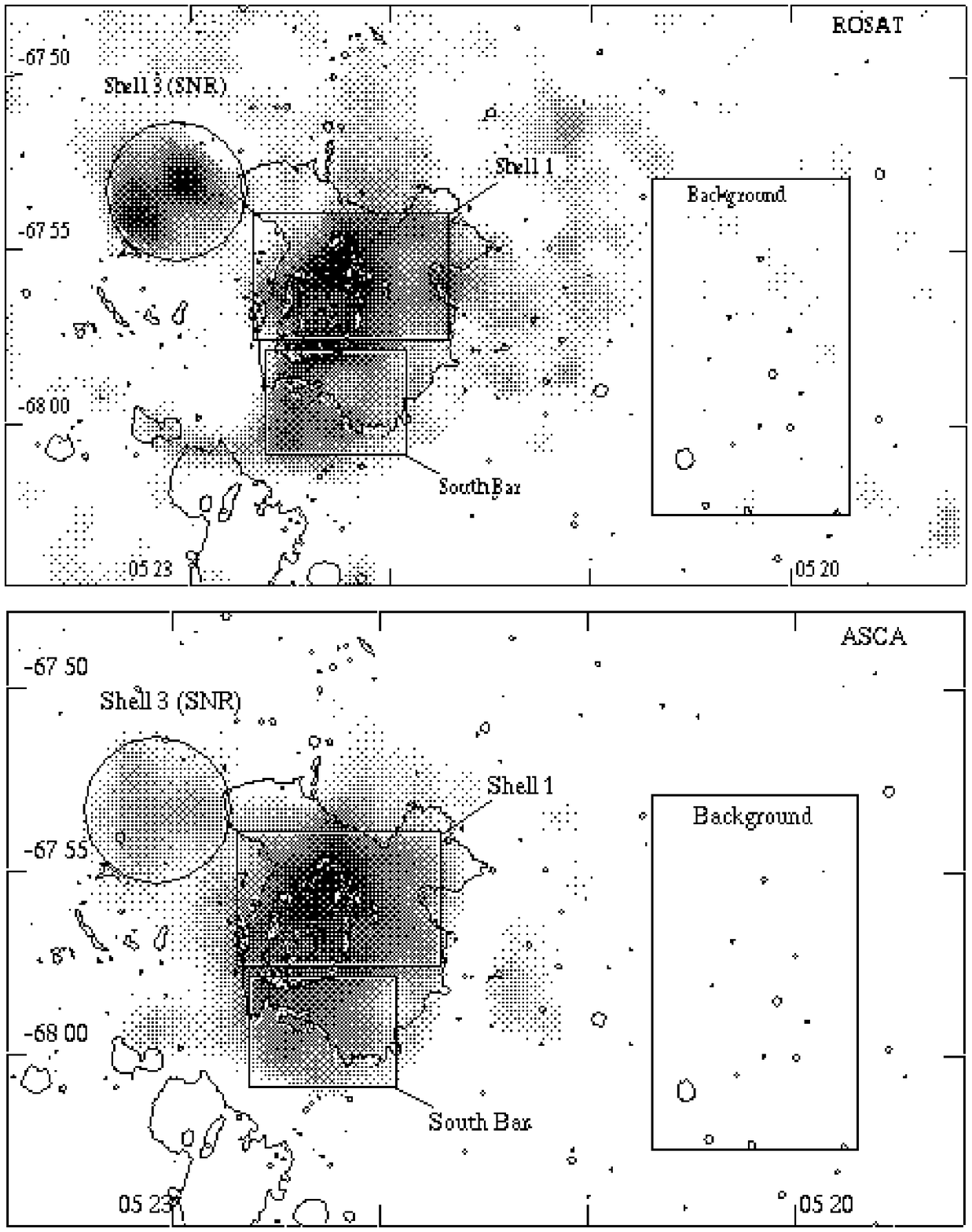,width=16.5cm,silent=}
\caption{\label{Xray}  }
\end{figure*}

\noindent
Figure~\ref{Xray}: top) ROSAT X-ray image, with \ha\ contours from our
CCD image overlaid.  The image has been smoothed with a Gaussian with
a FWHM of 45\arcsec.  Four of the features described by Chu \etal,
(1993) are labeled, and the regions from which spectra were extracted
are marked.  bottom) ASCA X-ray image, with \ha\ contours overlaid.
This image is at the same scale and orientation as (a), and has also
been smoothed with a Gaussian with a FWHM of 60\arcsec.  In both
figures, the contours are draw in pairs, with a white line at a
slightly lower value than the black line. This allows the reader to
distinguish positive and negative contours.

\begin{figure*}
\centerline{ \psfig{file=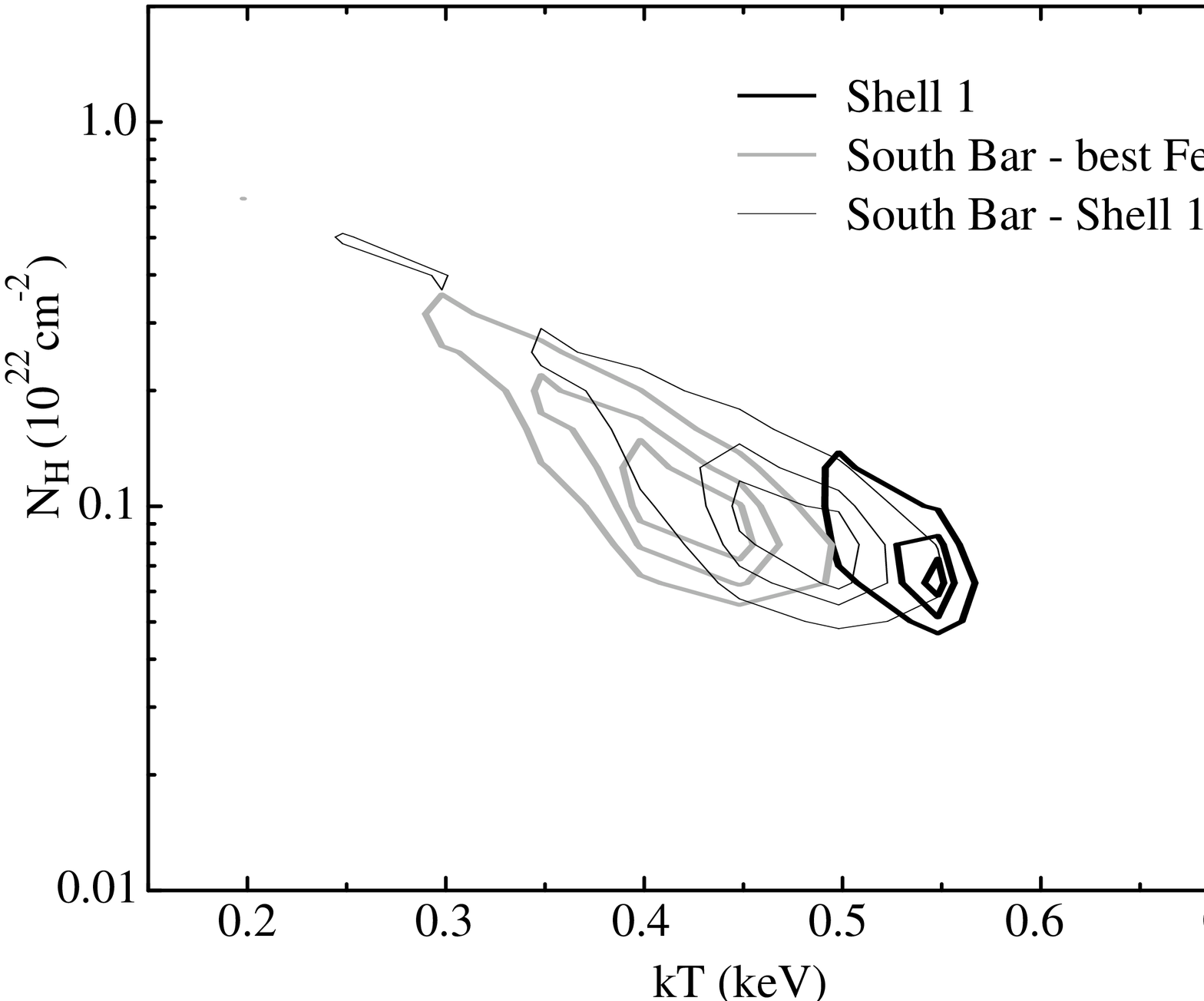,width=14cm,silent=}}

\begin{minipage}{8cm} \psfig{file=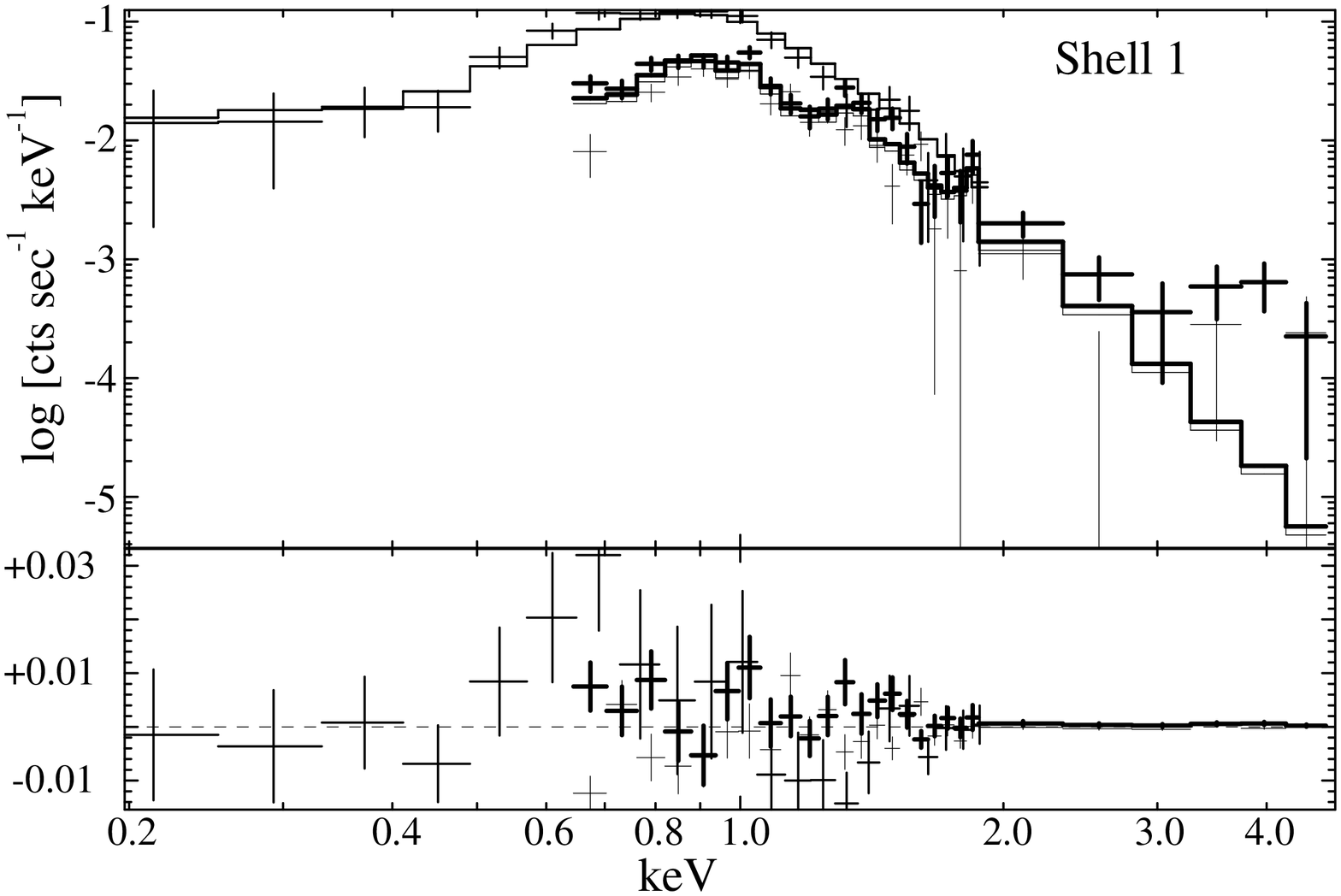,width=8cm,silent=}
\end{minipage}
\begin{minipage}{8cm} \psfig{file=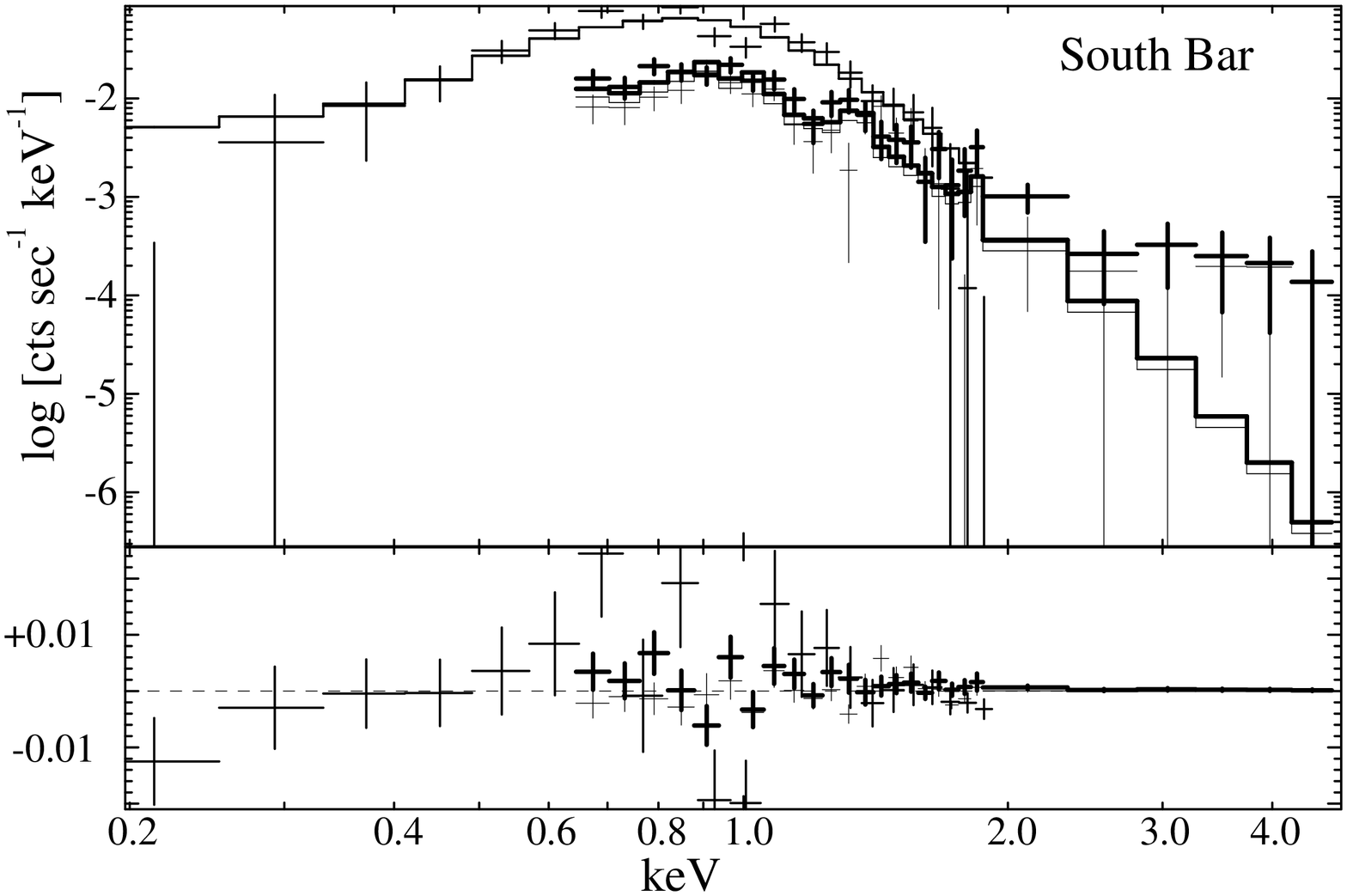,width=8cm,silent=}
\end{minipage}
\caption{\label{S1SB} }
\end{figure*}

\noindent
Figure~\ref{S1SB}: Comparison of Shell 1 and South Bar.  top)
Confidence contours for the spectral fits for Shell~1 (thick, black
line) and the South Bar (thick, gray line) in which the Fe abundance
is fixed at the best fit value.  Also plotted is the confidence
contours for the spectral fit for the South Bar, using the best-fit Fe
abundance value for Shell~1.  The three sets of contours represent
68\%, 90\%, and 99\% confidence levels.  bottom) Observed spectra
(error bars) and MEKA fits (histograms) for Shell 1 (left) and the
South Bar (right).  In both of the figures, we have included data from
ASCA SIS 0 (thick line), SIS 1 (thin line), and ROSAT PSPC (medium
weight line).

\begin{figure}
\psfig{file=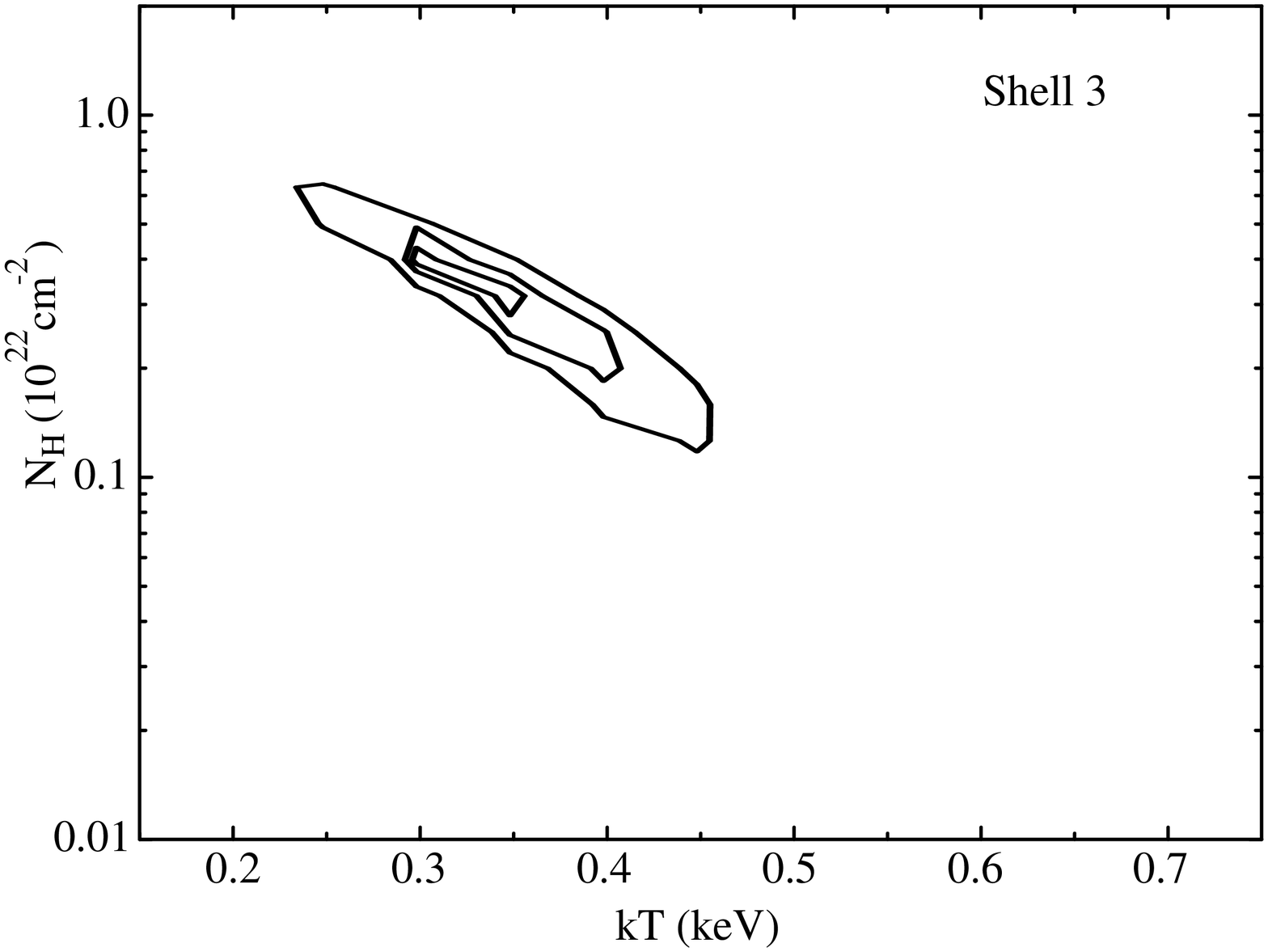,width=8cm,silent=}
\psfig{file=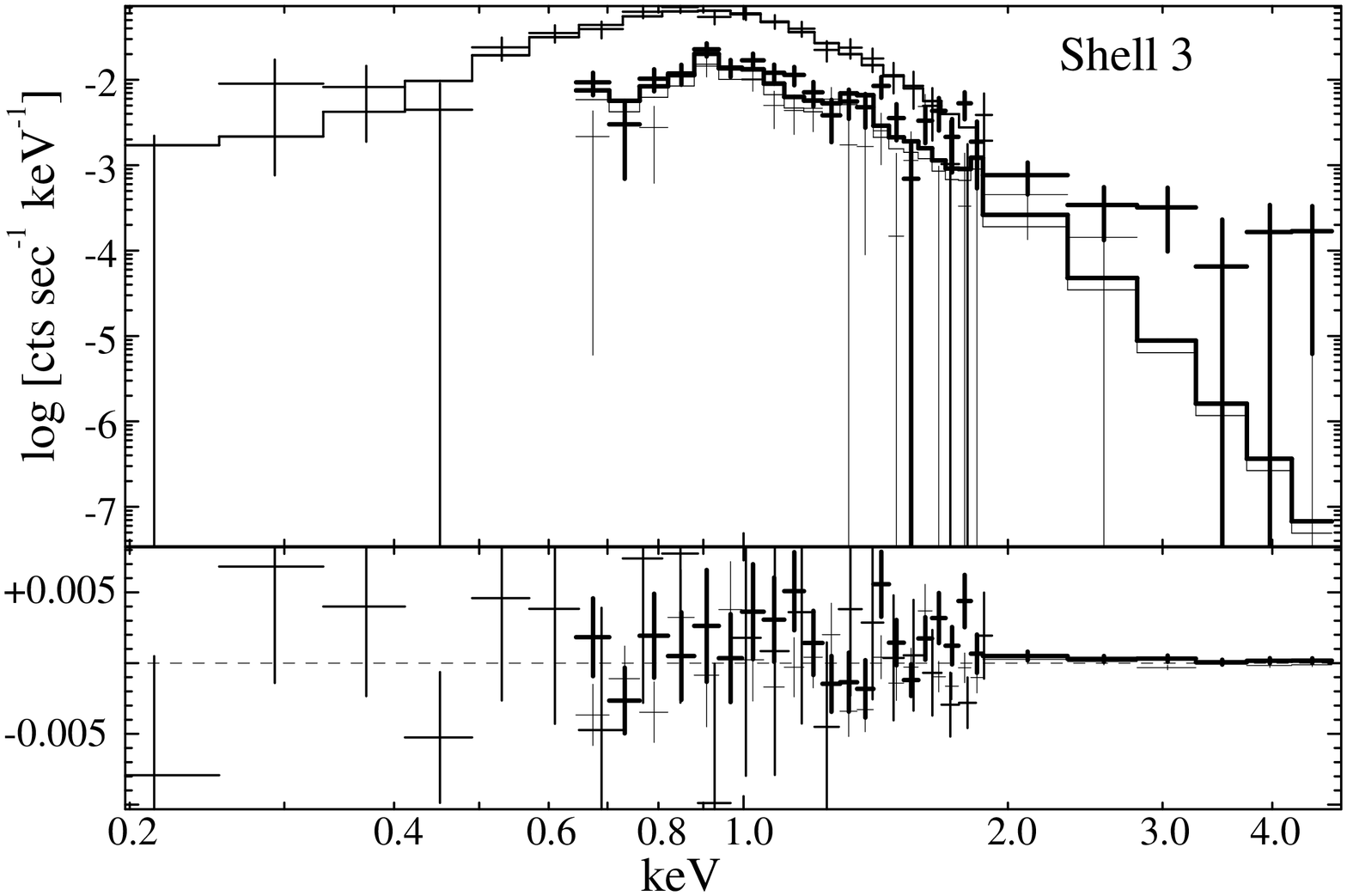,width=8cm,silent=}
\caption{\label{Shell3} }
\end{figure}

\noindent
Figure~\ref{Shell3}: left) Confidence contours for the spectral fits
for Shell~3 in which the Fe abundance is fixed at the best fit value.
The three sets of contours represent 68\%, 90\%, and 99\% confidence
levels.  right) Observed spectra (error bars) and MEKA fits
(histograms) for Shell 3.  In this figures, we have included data from
ASCA SIS 0 (thick line), SIS 1 (thin line), and ROSAT PSPC (medium
weight line).

\newpage

\end{document}